\documentclass[aoas,MSNbibl,nameyear,seceqn,dvips]{arximspdf}
\usepackage{graphicx}
\doi{10.1214/13-AOAS706} 
\volume{8}
\issue{2}
\pubyear{2014}
\firstpage{886}
\lastpage{904}

\makeatletter
\newcommand{\rrvert}{\vert}
\newcommand{\llvert}{\vert}
\newcommand{\overset}{\stackrel}
\makeatother

\begin{document}
\begin{frontmatter}

\title{Estimation of nonlinear differential equation model for
glucose--insulin dynamics in type I diabetic patients using generalized
smoothing\thanksref{T1}}
\runtitle{Optimized generalized smoothing of nonlinear ODE models}

\begin{aug}
\author[A]{\fnms{Inna} \snm{Chervoneva}\corref{}\thanksref{m1}\ead[label=e1]{Inna.Chervoneva@jefferson.edu}},
\author[A]{\fnms{Boris} \snm{Freydin}\thanksref{m1}\ead[label=e2]{Boris.Freydin@jefferson.edu}},
\author[B]{\fnms{Brian} \snm{Hipszer}\thanksref{m1}\ead[label=e3]{brian.hipszer@gmail.com}},\\
\author[C]{\fnms{Tatiyana V.} \snm{Apanasovich}\thanksref{m2}\ead[label=e4]{apanasovich@gwu.edu}}
\and
\author[B]{\fnms{Jeffrey I.} \snm{Joseph}\thanksref{m1}\ead[label=e5]{Jeffrey.Joseph@jefferson.edu}}
\runauthor{I. Chervoneva et al.}
\affiliation{Thomas Jefferson University\thanksmark{m1} and George
Washington University\thanksmark{m2}}
\address[A]{I. Chervoneva\\
B. Freydin\\
Division of Biostatistics\\
Thomas Jefferson University\\
Philadelphia, Pennsylvania 19107\\
USA\\
\printead{e1}\\
\phantom{E-mail:\ }\printead*{e2}} 
\address[B]{B. Hipszer\\
J. I. Joseph\\
Department of Anesthesiology\\
Thomas Jefferson University\\
Philadelphia, Pennsylvania 19107\\
USA\\
\printead{e3}\\
\phantom{E-mail:\ }\printead*{e5}}
\address[C]{T. V. Apanasovich\\
Department of Statistics\\
George Washington University\\
Washington, DC 20052\\
USA\\
\printead{e4}}
\end{aug}
\thankstext{T1}{Supported in part by Grant DK081088 from the NIH/NIDDK.}

\received{\smonth{10} \syear{2012}}
\revised{\smonth{8} \syear{2013}}

%
\begin{abstract}
In this work we develop an ordinary differential equations (ODE) model
of physiological regulation of glycemia in type 1 diabetes mellitus
(T1DM) patients in response to meals and intravenous insulin infusion.
Unlike for the majority of existing mathematical models of
glucose--insulin dynamics, parameters in our model are estimable from a
relatively small number of noisy observations of plasma glucose and
insulin concentrations. For estimation, we adopt the generalized
smoothing estimation of nonlinear dynamic systems of Ramsay et~al.
[\textit{J. R. Stat. Soc. Ser. B Stat. Methodol.} \textbf{69} (2007) 741--796].
In this framework, the ODE solution is approximated with a
penalized spline, where the ODE model is incorporated in the penalty.
We propose to optimize the generalized smoothing by using penalty
weights that minimize the covariance penalties criterion
(Efron [\textit{J. Amer. Statist. Assoc.} \textbf{99} (2004) 619--642]). The covariance penalties criterion provides an estimate of the
prediction error for nonlinear estimation rules resulting from
nonlinear and/or nonhomogeneous ODE models, such as our model of
glucose--insulin dynamics. We also propose to select the optimal number
and location of knots for B-spline bases used to represent the ODE
solution. The results of the small simulation study demonstrate
advantages of optimized generalized smoothing in terms of smaller
estimation errors for ODE parameters and smaller prediction errors for
solutions of differential equations. Using the proposed approach to
analyze the glucose and insulin concentration data in T1DM patients, we
obtained good approximation of global glucose--insulin dynamics and
physiologically meaningful parameter estimates.
\end{abstract}

%

\end{frontmatter}
\newpage

\section{Introduction}\label{sec1}

Diabetes mellitus is a chronic metabolic disease associated with
abnormalities in glucose metabolism that affect the uptake of glucose by
tissues and causes abnormal glucose excursions in the blood. Type 1 diabetes
mellitus (T1DM) is characterized by total insulin deficiency. In
insulin-deficient persons, the blood glucose levels have been roughly
controlled using insulin alone. Insulin can be injected (multiple daily
injection) or infused (continuous intravenous or subcutaneous insulin
infusion). Glucose levels of hospitalized diabetic patients are usually
managed with intravenous insulin infusions guided by glucose readings either
from a lab-based blood sample assay or point of care glucose meter. However,
infrequent testing of blood glucose levels does not provide sufficient trend
information to achieve the desired goals of near-normal glucose control.
Better glycemic control in T1DM in-hospital patients may be achieved
utilizing the continuous intravenous (IV) insulin infusion or insulin pump
delivering rapid-acting insulin continuously through a subcutaneous tissue
catheter, but automated regulation of the insulin delivery requires
constantly updated information about the blood glucose levels and
appropriate algorithms. Recently developed continuous glucose monitoring
sensors continuously measure the concentration of glucose in the blood or
interstitial fluid and display an averaged glucose value every one to five
minutes. However, there is no commonly accepted algorithm available for
controllers to allow automated regulation of the insulin delivery based on
the sensor feedback. Such an algorithm requires a relatively simple
parsimonious model for glucose--insulin dynamics validated using real patient
data. We propose a parsimonious model, which builds upon previously
considered models of glucose--insulin dynamics and may be estimated
using a
moderate number of plasma glucose and insulin measures and further used for
designing algorithms for an automated insulin delivery system.

\begin{figure} 

\includegraphics{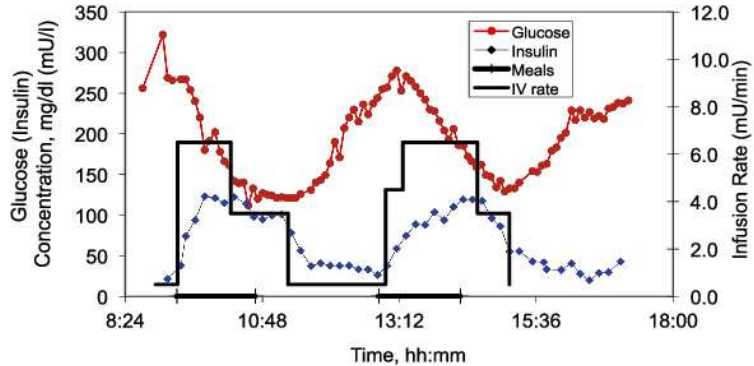}

\caption{Glucose concentration, insulin concentration and intravenous
insulin infusion rates and meal times as a function of study time in
one T1DM subject.}\label{Fig1}
\end{figure}

The motivating glucose--insulin dynamics data for four T1DM subjects were
collected as a part of the VIA Blood Glucose Monitor Study conducted at the
Jefferson Artificial Pancreas Center of Thomas Jefferson University,
Philadelphia, PA. The VIA Blood Glucose Monitor (currently GlucoScout,
International Biomedical, Austin, TX) is an FDA approved device that
can be
connected to an existing catheter inserted in a peripheral vein.
Samples of
blood are automatically transported from the bloodstream into the
flow-through sensor. The concentration of glucose is measured using an
enzyme-based electrochemical method. This self-calibrating device can
perform measurements as frequently as every five minutes. Blood glucose
samples were collected during an 8.5-hour protocol with breakfast at 30
minutes, lunch at 240 minutes and exercise at 450 minutes for 30 minutes.
Respectively, the first 7.5 hours (450 minutes) were used to fit the
proposed model of physiological regulation of glycemia in response to meals
and insulin infusion. The measurements collected included blood glucose
every 5 minutes with VIA Blood Glucose Analyzer and plasma insulin levels
every 10 minutes. The insulin infusion protocol provided for a 120-minute
square-wave bolus starting simultaneously with ingestion of the meal. The
initial dose was based upon a preprandial blood glucose measurement
(capillary blood sample tested on a commercial glucometer). Sixty minutes
after the beginning of the meal, the square-wave insulin bolus was adjusted
based on a second capillary blood measurement. The basal infusion rate (0.5
U/hr) of insulin was added to the bolus dose over the 120-minute meal
period. After 120 minutes, the bolus dose was completed, and the infusion
rate returned to the original basal level. Glucose and insulin
concentrations and IV insulin infusion rates as a function of study
time as
well as the times of the meals for one T1DM subject are shown in Figure~\ref{Fig1}.

%

Many mathematical models of glucose--insulin dynamics in humans and animals
have been proposed to date [e.g., \citet{Sheetal};
\citet{Beretal79}; \citet{Fisetal84};
\citet{Sor85}; \citet{CobBieFer90}; \citet{LehDeu92};
\citet{Traetal93}; \citet{Shietal97}; \citet{Wor97};
\citet{Leh}; \citet{DeGAri00}; \citet{Wiletal05}]. All existing models describe
glucose--insulin dynamics in terms of a system of nonlinear ordinary
differential equations (ODE). It is not feasible to convert most of these
models directly into estimable statistical models because of the large
number of latent (mostly not measurable) variables describing the
time-dependent levels of glucose, insulin, free fatty acids and
glucagon in
internal physiological compartments, such as heart, liver etc. The
validation of these models consist of finding a subset of parameters that
produce predicted glucose profiles closest to the observed data in a least
squares sense using an iterative curve-fitting algorithm and exhaustive grid
search. Other parameters are fixed at known biologically plausible levels.

In this work, a relatively simple model of glucose--insulin dynamics is
developed for physiological regulation of glycemia in response to meals and
IV insulin infusion in T1DM subjects. The model is validated using the real
data collected from four T1DM subjects studied in the clinical research
unit. To our knowledge, models of glucose dynamics in response to meals and
insulin infusion in T1DM subjects have not been reported in the literature
before. It is understood that the proposed model is a simplified
approximation of the true glucose--insulin dynamics. It is expected that
exact numeric solution of the estimated ODE model would capture global
dynamics of the physiological process, but not necessarily provide good fit
to observed data in all time subintervals. Therefore, we adapt the
generalized smoothing (generalized profiling) approach of \citet{Rametal07}, which is designed to handle data that may be
viewed as an approximate rather than precise solution of the postulated ODE
model. Section~\ref{sec3} provides a short formal description of the generalized
profiling approach. Some alternative smoothing-based methods of ODE model
estimation [e.g., \citet{Var82}; \citet{Bru08}; \citet{CheWu08}]
proceed with
first finding a nonparametric smoother of the data and then considering this
smoother to be a solution of the ODE model in order to estimate ODE
parameters. Nonlinear least squares [e.g., \citet{LiOsbPrv05};
\citet{XueMiaWu10}] and Bayesian methods [e.g., \citet{HuaLiuWu06};
\citet{DonSam07}] are other approaches for estimating
parameters of
the ODE models.

Implementation of the generalized smoothing approach requires selecting the
basis for representing the ODE solution and the penalty weight(s).
Here, we
propose to optimize the generalized smoothing solutions of nonlinear and
nonhomogeneous ODEs by selecting the penalty weight(s) so that the
estimated prediction error is minimized. As an unbiased estimate of the
prediction error, we adapt the covariance penalty [\citet{Efr04}], which is
suitable for nonlinear prediction rules. In the context of generalized
profiling, the penalty is used to express the difference between the
derivative of the ODE solution approximation and the right-hand side of the
ODE model
and does not serve the usual purpose of regularization. Therefore, for
regularization and performance optimization, it is proposed to use B-spline
bases with optimized number and location of the knots using the knot
selection methodology and software recently developed by \citet{Spietal13}.

In Section~\ref{sec2} we develop a parsimonious model of glucose--insulin
dynamics in
T1DM patients. Generalized profile estimation methodology is summarized in
Section~\ref{sec3}. Section~\ref{sec4} describes the proposed optimization for generalized
profile estimation of nonlinear and/or nonhomogeneous ODE models. The
optimized generalized profiling is applied to estimate the new model of
glucose--insulin dynamics using real data collected from T1DM patients in
Section~\ref{sec5}. A simulation study is presented in Section~\ref{sec6}. Section~\ref{sec7} concludes
with discussion.

\section{Model of glucose--insulin dynamics in T1DM patients}\label{sec2}

A comprehensive glucose--insulin dynamics model in T1DM patients may be
viewed as comprised of three components: a glucose metabolism model, a meal
model, and an insulin kinetics model. The model of glucose metabolism
describes the regulation of glucose uptake and production in the body---primarily controlled by the concentrations of glucose and insulin. The meal
model describes the rate of glucose absorption from the intestine. The model
of insulin kinetics is required to describe the rate at which insulin
diffuses into, and is eliminated from, the systemic circulation.

Most of the glucose--insulin dynamics models developed to date focus on
nondiabetic or type 2 diabetic patients, for whom the plasma glucose
concentration affects the insulin production. In T1DM patients, the
pancreas does not release insulin, and the model has to incorporate an
exogenous insulin input. Meanwhile, it is plausible to assume that plasma
insulin concentration affects the glucose concentration similarly in
diabetic and nondiabetic patients. Recently proposed models for glucose and
insulin dynamics after an IV glucose tolerance test (IVGTT) [\citet{DeGAri00};\vspace*{-2pt}
\citet{Mukhopadhyay.etal2004}] assume the following model
for the
derivative $\overset{\bullet}{G}(t)$ of plasma glucose concentration $G(t)$
[mg/dl] as a function of the plasma insulin concentration $I(t)$
[mU/l]:%
\begin{equation}
\overset{\bullet} {G}(t)=b_{0}-b_{1}G(t)-b_{2}G(t)I(t),
\label{IVGTT}
\end{equation}
where $b_{0}$ [(mg/dl)~min$^{-1}]$ is the constant increase in plasma glucose
concentration due to constant baseline liver glucose release, $b_{1}$
[min$^{-1}$] is the spontaneous glucose 1st order disappearance rate,
and $%
b_{2}$ [min$^{-1}$ (l/mU)] is the insulin-dependent glucose
disappearance rate. This model has to be extended to incorporate the terms
describing the external glucose absorption from the meals. In this
work, we
adapt one of the most parsimonious meal absorption models, which assumes
that the rate of glucose appearance in the blood follows the model $\mu
(t-t_{M})_{+}\exp ( \nu(t-t_{M})_{+} )$, where $t_{M}$ is the
starting time of the meal, $\nu$ is the parameter related to the rate of
glucose absorption from the gut and $\mu$ is the product of the
distribution volume and the constant related to the total amount of
carbohydrates consumed [\citet{Mon}]. With IV insulin infusion, the
kinetics of plasma insulin concentration $I(t)$ [mU/l] is described
by the\vspace*{-1pt}
one-compartment model [e.g., \citet{Steetal03}; \citet{HipJosKam05}]
$\overset{\bullet}{I}(t)=-c_{1}I(t)+c_{2}r(t)$,
where $c_{1}$ [min$^{-1}$] is the insulin 1st order disappearance rate, $%
c_{2}$ [l$^{-1}]$ is the reciprocal of the volume of the insulin
distribution space, and $r(t)$ [mU/min] is the IV insulin\vspace*{-1pt} infusion rate.
For nondiabetic and type 2 diabetic patients, $\overset{\bullet}{I}(t)$
also depends on the glucose concentration $G(t)$, but for T1DM patients,
the pancreas does not release insulin, and insulin concentration $I(t)$ is
independent of $G(t)$. Thus, we consider the following nonhomogeneous
nonlinear ODE model for an individual T1DM subject on IV insulin delivery
and consuming $L$ meals at times $t_{M_{1}},\ldots, t_{M_{L}}$:%
\begin{eqnarray}
\overset{\bullet} {G}(t)&=&b_{0}-b_{1}G(t)-b_{2}G(t)I(t)
\nonumber\\[-8pt] \label{mainODE}  \\[-8pt]
&&{} + \sum_{i=1}^{L}\mu _{i}(t-t_{M_{i}})_{+}
\exp \bigl( \nu_{i}(t-t_{M_{i}})_{+} \bigr),\nonumber
\\
\label{Ht} \overset{\bullet} {I}(t)&=&-c_{1}I(t)+c_{2}r(t).
\end{eqnarray}

For nonlinear ODE systems, identifiability of all parameters is not a
trivial issue. In our models, the second differential equation (\ref{Ht})
is independent of $G(t)$, linear and has a closed-form solution. Hence,
parameters $c_{1}$ and $c_{2}$ are identifiable based on measured
$I(t)$ and
known $r(t)$. Identifiability of parameters in (\ref{mainODE}) is
straightforward to show using the multiple time points method of \citet{Wuetal08}.

\section{Generalized profiling estimation of ODE models}\label{sec3}

Following \citet{Rametal07}, it is assumed that a vector of $d$ output
functions or states $\mathbf{x}(t)= [ x_{1}(t),\ldots,x_{d}(t) ],
t\in[0,T]$ depends on a vector of $g $ input functions $\mathbf
{u}(t)= [ u_{1}(t),\ldots,u_{g}(t) ] $. The ODE model is written
for the
derivative vector $\overset{\bullet}{\mathbf{x}}(t)$ as a known vector
function $\mathbf{f}$ of $t, \mathbf{x}(t)$, $\mathbf{u}(t)$, and
unknown $%
p\times1$ parameter vector $\bolds{\theta}$,
\begin{equation}
\overset{\bullet} {\mathbf{x}}(t)=\mathbf{f} \bigl( \mathbf{x}(t),\mathbf
{u}(t),t,\bolds{\theta} \bigr),\qquad t\in [0,T]. \label{ssODE}
\end{equation}
It is assumed that a subset of $m\leq d$ output functions is measured with
error on some grid of points $0\leq t_{jl}\leq T, j\in\mathcal
{I}\subset
\{1,\ldots,d\}, 1\leq l\leq N_{j}$ (possibly different grids for
different $%
j $). The measurement error model is $y_{jl}=x_{j}(t_{jl})+\varepsilon
_{jl}$, where $y_{jl}=y ( t_{jl} ) $ are observed values and $
\varepsilon_{jl}\sim \mbox{i.i.d. } N(0,\sigma_{j}^{2})$. Each state
$x_{j}(t), j=1,\ldots,d$, is approximated by an expansion with respect to some basis $
B_{j}= \{ \psi_{jk}(t),1\leq k\leq K_{j} \}$,
\begin{equation}
\tilde{x}_{j}(t)=\sum_{k=1}^{K_{j}}
\alpha_{jk}\psi_{jk}(t)=\bolds {\alpha}_{j}^{\prime}
\bolds{\psi}_{j}(t)=\bolds{\psi}_{j}^{\prime
}(t)%
\bolds{\alpha}_{j}, \label{predY}
\end{equation}
where $\bolds{\alpha}_{j}= [ \alpha_{j1},\ldots,\alpha
_{jK_{j}} ]
^{\prime}$ and $\bolds{\psi}_{j}(t)= [ \psi_{j1}(t),\ldots,\psi
_{jK_{j}}(t) ] ^{\prime}$.

The data fitting criterion is defined as a negative log-likelihood
\begin{equation}
\mathbf{H}(\bolds{\sigma},\bolds{\alpha})=-\sum_{j=1}^{m}
\ln \bigl\{ g ( \mathbf{e}_{j}|\sigma_{j},\bolds{
\alpha}_{j} ) \bigr\} \label{Hdtf}
\end{equation}
of the error terms vectors $\mathbf{e}_{j}=\mathbf{y}_{j}-\bolds{\Psi}_{j}\bolds{\alpha}_{j}$, where $\mathbf{y}_{j}= [
y_{j1},\ldots,y_{jN_{j}} ]$, $\bolds{\Psi}_{j}$ is the
$N_{j}\times K_{j}$ matrix of basis
functions $\psi_{j1}(t),\ldots,\psi_{jK_{j}}(t)$ evaluated at $N_{j}$ time
points, $\bolds{\alpha}= [ \bolds{\alpha}_{1}^{\prime
},\ldots,\bolds{\alpha}_{m}^{\prime} ] ^{\prime}, \bolds
{\sigma}= [ \sigma
_{1},\ldots,\sigma_{m} ] ^{\prime}$, and $g ( \mathbf
{e}_{j}|\sigma
_{j},\bolds{\alpha}_{j} ) $ is the density of the error terms. The
estimate $\hat{\bolds{\alpha}}$ of the parameter $\bolds{\alpha}$ is
computed by minimizing the penalized criterion%
\begin{equation}
\mathbf{J}(\bolds{\alpha},\bolds{\theta},\bolds{\lambda})=\mathbf {H}(\bolds{\sigma},
\bolds{\alpha})+\sum_{j=1}^{d}\lambda
_{j}\mathrm{PEN}_{j}(\tilde{\mathbf{x}}|\bolds{\theta}),
\label{penJ}
\end{equation}
where $\bolds{\lambda}= [ \lambda_{1},\ldots,\lambda_{d} ]
^{\prime}$ and the penalty for the $j$th output function is defined
by%
\begin{equation}
\mathrm{PEN}_{j}(\tilde{\mathbf{x}}|\bolds{\theta})=\int \biggl\{
\frac
{d}{dt}\tilde{x}_{j}(t)-f_{j} \bigl(
\tilde{\mathbf {x}}(t),\mathbf{u}(t),t,%
\bolds{\theta} \bigr)
\biggr\} ^{2}\,dt. \label{PEN}
\end{equation}
The composite penalty $\sum_{j=1}^{m}\lambda_{j}\mathrm{PEN}_{j}(\tilde
{\mathbf{x%
}}|\bolds{\theta})$ measures the closeness of $\tilde{\mathbf{x}}(t)$
to the solution of (\ref{ssODE}). Parameter $\bolds{\alpha}$ is
considered a nuisance parameter, which depends on the structural
parameter $%
\bolds{\theta}$ and smoothing parameter $\bolds{\lambda}$, with function
$\bolds{\alpha}=\bolds{\alpha} ( \bolds{\theta},\bolds
{\lambda} ) $ defined implicitly (under the assumptions of the
implicit function
theorem) by%
\begin{equation}
\frac{\partial\mathbf{J}(\bolds{\alpha},\bolds{\theta},\bolds{\lambda})}{\partial\bolds{\alpha}}=\mathbf{0} \label{impFun}
\end{equation}
in some open neighborhood of the minimum of the penalized criterion
(\ref%
{penJ}). The generalized profiling approach (a.k.a. generalized smoothing)
estimates the structural parameter vector $\bolds{\theta}$ by
minimizing (\ref%
{Hdtf}) with respect to $\bolds{\theta}$ using the Gauss--Newton
algorithm. The
necessary gradient vector is
\[
\frac{d\mathbf{H}(\bolds{\alpha} ( \bolds{\theta} )
)}{d\bolds{\theta}^{\prime}}=\frac{\partial\mathbf{H}(\bolds
{\alpha} ( \bolds{\theta} ) )}{\partial\bolds{\alpha}^{\prime}}\frac{d\bolds{\alpha} ( \bolds{\theta} ) }{d\bolds{\theta}^{\prime}},
\]
where $d\bolds{\alpha} ( \bolds{\theta},\bolds{\lambda} )/d
\bolds{\theta}^{\prime}$ is computed using the implicit function theorem,
\begin{equation}
\frac{d\bolds{\alpha} ( \bolds{\theta} ) }{d\bolds{\theta}^{\prime}}=- \biggl( \frac{\partial^{2}\mathbf{J}(\bolds{\alpha},\bolds{\theta},\bolds{\lambda})}{\partial\bolds{\alpha}^{\prime}\,\partial\bolds
{\alpha}} \biggr) ^{-1}
\frac{\partial^{2}\mathbf{J}(\bolds{\alpha},\bolds{\theta},\bolds{\lambda})}{d\bolds{\theta}^{\prime}\,\partial\bolds{\alpha}}. \label{dadg}
\end{equation}
After each update of $\bolds{\theta}$, parameter $\bolds{\alpha}$ is
updated by minimizing (\ref{penJ}) conditionally on the current value
of $%
\bolds{\theta}$ and a priori chosen $\bolds{\lambda}$.
Representation of
$\tilde{x}_{j}(t)$ through the basis expansion (\ref{predY})
automatically yields the derivative needed to evaluate (\ref{PEN}),
\[
\frac{d}{dt}\tilde{x}_{j}(t)=\sum
_{k_{0}}^{K_{j}}\alpha_{jk}\frac
{d}{dt}
\psi_{jk}(t)=\bolds{\alpha}_{j}^{\prime}\overset{
\bullet } {\bolds{\psi}}_{j}(t)=\overset{\bullet} {\bolds{\psi
}}_{j}(t)^{\prime}\bolds{\alpha}_{j}.
\]
Notably, the penalized criterion (\ref{penJ}) is not minimized jointly with
respect to~$\bolds{\theta}$~and~$\bolds{\alpha}$ because joint
minimization yielded unsatisfactory estimates [\citet{HecRam00}].
When $e_{jl}\sim \mbox{i.i.d. }N(0,\sigma_{j}^{2})$, minimizing (\ref{Hdtf}) is
equivalent to minimizing%
\begin{equation}
\mathbf{H}(\bolds{\alpha},\bolds{\sigma})=\sum_{j=1}^{m}w_{j}
( \mathbf {y}_{j}-\bolds{\Psi}_{j}\bolds{
\alpha}_{j} ) ^{\prime} ( \mathbf{y%
}_{j}-
\bolds{\Psi}_{j}\bolds{\alpha}_{j} ), \label{Hdtf2}
\end{equation}
where\vspace*{-1pt} $w_{j}=\sigma_{j}^{-2}$ [\citet{Rametal07} mention other
alternatives for $w_{j}$]. We use~(\ref{Hdtf2}) with weights
$w_{j}=\sigma
_{j}^{-2} $ estimated prior to the generalized profile estimation along with
the starting values and drop dependence of $\mathbf{H}$ on $\bolds
{\sigma}$
from further notation.

In the original profiled estimation work, \citet{Rametal07} assume
that $%
\bolds{\lambda}$ is fixed a priori. \citet{CaoRam09} view
$\bolds{\lambda}$ as a complexity parameter and propose optimization
with respect
to $\bolds{\lambda}$ as the third outer level of optimization.
Denoting a
suitable outer optimization criterion by $\mathbf{F}(\bolds{\lambda},\bolds{\theta}
(\bolds{\lambda}),\bolds{\alpha} ( \bolds{\theta},\bolds{\lambda} ) )$, they propose
to optimize $\mathbf{F}(\bolds{\lambda},\bolds{\theta}(\bolds{\lambda}),\bolds{\alpha} (
\bolds{\theta},\bolds{\lambda} ) )$ as a function of $\bolds{\lambda}$, so
that every time $\bolds{\lambda}$ is changed, $\bolds{\theta}$
is updated by minimizing $\mathbf{H}(\bolds{\alpha} ( \bolds{\theta},
\bolds{\lambda} ) )$ as described above. Respectively, for every
change in $\bolds{\theta}$, the estimates of the nuisance parameters $
\bolds{\alpha} ( \bolds{\theta},\bolds{\lambda} ) $ are updated by
minimizing $\mathbf{J}(\bolds{\alpha},\bolds{\theta},\bolds{\lambda})$. Thus,
the overall (outer) optimization is carried on with respect to $\bolds{\lambda}$, the middle level of optimization of $\mathbf{H}(
\bolds{\alpha} ( \bolds{\theta},\bolds{\lambda} ) )$ is nested
within the
outer level optimization, and the inner level of optimization of
$\mathbf{J}(%
\bolds{\alpha},\bolds{\theta},\bolds{\lambda})$ is nested within the middle
level optimization. The term ``parameter cascade'' was introduced to reflect
this multistage optimization with respect to $\mathbf{F}$, $\mathbf{H}$ and
$\mathbf{J}$.

\section{Optimization of generalized profiling estimation}\label{sec4}

For generalized\break  smoothing, the most important question is the choice of the
penalty weights $\bolds{\lambda}$. \citet{CaoRam09} adapted
the generalized cross-validation (GCV) criterion [\citet{CraWah78}] to
select the weights for the penalties that may be represented as a quadratic
form in the spline coefficients, which corresponds to linear differential
operators defined by the ODE model. For the models with nonlinear and/or
nonhomogeneous ODEs, this approach is not appropriate because there is no
associated smoothing matrix that would linearly transform data into spline
coefficients. On the other hand, a direct implementation of leave-one-out
cross-validation to minimize the prediction error would be extremely
computationally intense for nonsparse functional data and nonlinear ODE
models. The GCV criterion provides an estimate of the prediction error for
linear estimation rules. In particular, the GCV criterion is an unbiased
estimate of the prediction error for penalized splines [e.g., \citet{Gu2002}].
Optimization of the tuning parameter $\bolds{\lambda}= [ \lambda
_{1},\ldots,\lambda_{m} ] ^{\prime}$ for the generalized smoothing of
nonlinear and/or nonhomogeneous ODEs may be based on an estimate of the
prediction error, which would be suitable for nonlinear estimation
rules as
well as for the linear ones. The covariance penalties\vadjust{\goodbreak} approach [\citet{Efr04}]
provides such an estimate of the prediction error. Following \citet{Efr04},
let us denote the mean function by $\bolds{\mu}$ and corresponding
prediction by $\hat{\bolds{\mu}}=m ( \mathbf{y} ) $,
where $%
m ( \mathbf{y} ) $ is a nonlinear estimation rule. An unbiased
estimate of the prediction error is given by
\[
\widehat{\mathrm{Err}}=\sum_{j=1}^{m}w_{j}
\biggl\{ \sum_{1\leq l\leq N_{j}} ( y_{jl}-\hat{
\mu}_{jl} ) ^{2}+2\sum_{1\leq l\leq
N_{j}}\operatorname{cov}(
\hat{\mu}_{jl},y_{jl}) \biggr\}.
\]
\citet{Efr04} used the parametric bootstrap to estimate $2\sum_{1\leq
l\leq
N_{j}}\operatorname{cov}(\hat{\mu}_{jl},y_{jl})$ (covariance penalty) in the general
case. In the case of Gaussian errors and a differentiable mapping
$\hat{\bolds{\mu}}=m ( \mathbf{y} ) $, $\operatorname{cov}(\hat{\mu
}_{jl},y_{jl})$
may be estimated by $\sigma_{j}^{2} ( \partial\hat{\mu
}_{jl}/\partial y_{jl} ) $ [\citet{Ste81}; \citet{Ye98}; \citet{Efr04}], so that
with $w_{j}=\sigma_{j}^{-2}$, one obtains
\[
\widehat{\mathrm{Err}}=\sum_{j=1}^{m} \biggl\{ \sum
_{1\leq l\leq N_{j}}w_{j} ( y_{jl}-\hat{
\mu}_{jl} ) ^{2}+2\sum_{1\leq l\leq N_{j}}
\frac
{\partial\hat{\mu}_{jl}}{\partial y_{jl}} \biggr\}.
\]
For linear prediction rules, using the covariance penalties is
asymptotically equivalent to using the GCV criterion [\citet{Efr04}].

If $ \{ \hat{\bolds{\theta}},\hat{\bolds{\alpha}} (
\hat{\bolds{\theta}},\bolds{\lambda} )
\} $ is the generalized smoothing solution computed for fixed $%
\bolds{\lambda}$, then the predicted mean function for state
$x_{j}(t)$ is
$\hat{\bolds{\mu}}_{j}=\bolds{\Psi}_{j}\hat{\bolds{\alpha}}_{j}$. Respectively, $\hat{\mu}_{jl}=\bolds{\psi
}_{j}(t_{jl})^{\prime
}\hat{\bolds{\alpha}}_{j}$ and (\ref{Hdtf2}) implies that $%
\sum_{j=1}^{m}\sum_{1\leq l\leq N_{j}}w_{j} ( y_{jl}-\hat{\mu
}_{jl} ) ^{2}=\mathbf{H} ( \hat{\bolds{\alpha}} (
\hat{\bolds{\theta}},\bolds{\lambda})  )$.
Further, $%
\frac{\partial}{\partial y_{jl}}\bolds{\psi}_{j}(t_{jl})^{\prime
}\hat{\bolds{\alpha}}_{j}=\bolds{\psi}_{j}(t_{jl})^{\prime
}\frac{\partial}{\partial y_{jl}}\hat{\bolds{\alpha}}_{j}$ and the
covariance penalties criterion for the generalized profiling may be written
as
\[
\mathbf{F} ( \bolds{\lambda},\hat{\bolds{\theta}},%
\hat{\bolds{\alpha}} ) =\mathbf{H} \bigl( \hat{\bolds{\alpha}} (
\hat{\bolds{\theta}},\bolds{\lambda} ) \bigr) +4\sum
_{j=1}^{m}\sum_{1\leq l\leq N_{j}}
\bolds{\psi }_{j}(t_{jl})^{\prime}\frac{\partial}{\partial y_{jl}}
\hat{\bolds{\alpha}}_{j}.
\]
Since equation $\partial\mathbf{J}(\bolds{\alpha},\bolds{\theta},\bolds{\lambda
})/\partial\bolds{\alpha}=\mathbf{0 }$ is satisfied by $\hat
{\bolds{\alpha}}$, the derivatives $\partial\hat{\bolds{\alpha}}_{j}/\partial y_{jl}$ may be computed using the inverse function
theorem,
\[
\frac{\partial\hat{\bolds{\alpha}}_{j}}{\partial y_{jl}}= - \biggl( \frac{\partial^{2}\mathbf{J}(\bolds{\alpha},\bolds{\theta},
\bolds{\lambda})}{\partial\bolds{\alpha}_{j}^{\prime}\,\partial\bolds{\alpha}_{j}} \biggr) ^{-1}
\frac{\partial^{2}\mathbf{J}(\bolds{\alpha},\bolds{\theta},%
\bolds{\lambda})}{\partial y_{jl}\,\partial\bolds{\alpha}_{j}} \biggl\vert _{\bolds{\alpha}=\hat{\bolds{\alpha}}},
\]
where $\partial^{2}\mathbf{J}(\bolds{\alpha},\bolds{\theta},\bolds{\lambda})/ \partial y_{jl}\,\partial\bolds{\alpha}_{j}=-2w_{j}\bolds{\psi
}_{j}(t_{jl})$ and expression for $\partial^{2}\mathbf{J}(\bolds{\alpha},\bolds{\theta},\bolds{\lambda})/\break \partial\bolds{\alpha}_{j}^{\prime
}\,\partial\bolds{\alpha}_{j}$ is given in the \hyperref[app]{Appendix}. Hence, the covariance
penalties criterion is
\begin{eqnarray}\label{outerF}
&& \mathbf{F}(\bolds{\lambda},\hat{\bolds{\theta}},\hat{\bolds {
\alpha}})
\nonumber\\[-8pt]\\[-8pt]
&&\qquad =\mathbf{H}(\hat{\bolds{\alpha}})+4\sum
_{j=1}^{m}w_{j}\sum_{1\leq l\leq N_{j}} \bolds{\psi }_{j}(t_{jl})^{\prime}
\biggl( \frac{\partial^{2}\mathbf{J}(\bolds{\alpha},\bolds{\theta},\bolds{\lambda})}{\partial\bolds{\alpha}_{j}^{\prime
}\,\partial\bolds{\alpha}_{j}} \biggr) ^{-1} \Biggl\vert \mathop{_{\bolds{\theta}=\hat{\bolds{\theta}}}}_{\bolds{\alpha}=\hat{\bolds{\alpha}}}\bolds{\psi }_{j}(t_{jl}).\nonumber\hspace*{-30pt}
\end{eqnarray}

It is proposed to compute $\bolds{\lambda}=\arg\min_{\bolds{\lambda}} \mathbf{F}  (\bolds{\lambda},\hat{\bolds{\theta}},
\hat{\bolds{\alpha}} ) $ using the Gauss--Newton
optimization to
find a point of minimum, such that at every iteration step, the profiled
solution $ \{ \hat{\bolds{\theta}},\hat{\bolds{\alpha}} ( \hat{\bolds{\theta}}, \bolds{\lambda
} )  \} $ is updated conditionally on $\bolds{\lambda}$
as described in Section~\ref{sec3}.\vadjust{\goodbreak}

The Gauss--Newton optimization requires the gradient $\frac{d\mathbf{F}(%
\bolds{\lambda},\hat{\bolds{\theta}},\hat{\bolds{\alpha}})}{d\bolds{\lambda}^{\prime}}$. Assuming that
differentiation under the integral is appropriate, this gradient is computed
as
\[
\frac{d}{d\bolds{\lambda}^{\prime}}\mathbf{F} ( \bolds{\lambda},\hat{\bolds{\theta}},\hat{\bolds{\alpha}} ) =\frac{\partial\mathbf{F}}{\partial\bolds{\lambda}^{\prime}}+
\frac{\partial\mathbf{F}}{\partial\bolds{\alpha}^{\prime}}
\frac
{d\bolds{\alpha}}{d\bolds{\lambda}^{\prime}} \biggl\vert_{\bolds{\theta}=\hat{\bolds{\theta}},\bolds{\alpha}=%
\hat{\bolds{\alpha}}},
\]
where $d\bolds{\alpha}/d\bolds{\lambda}^{\prime}$ may be
computed the same way as $d\bolds{\alpha}/d\bolds{\theta}^{\prime
}$ in (\ref{dadg}),
\[
\frac{\partial\bolds{\alpha}}{\partial\bolds{\lambda}^{\prime
}}=- \biggl( \frac{\partial^{2}\mathbf{J}(\bolds{\alpha},\bolds{\theta
},\bolds{\lambda})}{\partial\bolds{\alpha}^{\prime}\,\partial
\bolds{\alpha}} \biggr) ^{-1}
\frac{\partial^{2}\mathbf{J}(\bolds{\alpha},\bolds{\theta},\bolds{\lambda})}{\partial\bolds{\lambda}^{\prime
}\,\partial\bolds{\alpha}},
\]
with $ \partial^{2}\mathbf{J}(\bolds{\alpha},\bolds{\theta},\bolds{\lambda})/\partial\lambda_{k}\,\partial\bolds{\alpha}^{\prime}$ given in the
\hyperref[app]{Appendix}, and
\begin{eqnarray*}
\frac{\partial\mathbf{F}}{\partial\lambda_{k}} &=&\sum_{j=1}^{m}4w_{j}
\sum_{1\leq l\leq N_{j}}\frac{\partial}{\partial
\lambda_{k}}\operatorname{trace} \biggl( \biggl(
\frac{\partial^{2}\mathbf{J}(\bolds{\alpha},\hat
{\bolds{\theta}},\bolds{\lambda})}{\partial\bolds{\alpha}_{j}^{\prime
}\,\partial\bolds{\alpha}_{j}} \biggr) ^{-1} \bigg\vert_{\bolds{\alpha}=%
\hat{\bolds{\alpha}}}\bolds{\psi}_{j}(t_{jl})\bolds{\psi }_{j}(t_{jl})^{\prime}
\biggr)
\\
&=&\sum_{j=1}^{m}4w_{j}\sum
_{1\leq l\leq N_{j}} \biggl[ \operatorname{Vec} \biggl\{ \biggl( \frac{\partial^{3}\mathbf{J}(\bolds{\alpha},\hat{\bolds{\theta}},\bolds{\lambda})}{\partial\lambda_{k}\,\partial
\bolds{\alpha}_{j}^{\prime}\,\partial\bolds{\alpha}_{j}}
\biggr) ^{-T} \bigg\vert_{\bolds{\alpha}=\hat{\bolds{\alpha}}} \biggr\} ^{T}\operatorname{Vec} \bigl(
\bolds{\psi}_{j}(t_{jl})\bolds{\psi }_{j}(t_{jl})^{\prime}
\bigr) \biggr],
\\
\frac{\partial\mathbf{F}}{\partial\bolds{\alpha}_{k}^{\prime}} &=&-2w_{k} ( \mathbf{y}_{k}-\bolds{\Psi}_{k}\bolds{\alpha}_{k} ) ^{\prime}
\bolds{\Psi}_{k}
\\
&&{}+\sum_{j=1}^{m}4w_{j}\sum
_{1\leq l\leq N_{j}}\frac{\partial}{\partial
\bolds{\alpha}_{k}^{\prime}} \biggl[ \operatorname{Vec} \biggl\{ \biggl(
\frac{\partial^{2}\mathbf{J}(\bolds{\alpha},\hat{\bolds{\theta}},\bolds{\lambda})}{\partial\bolds{\alpha}_{j}^{\prime}\,\partial
\bolds{\alpha}_{j}} \biggr) ^{-T} \bigg\vert_{\bolds{\alpha}=\hat{\bolds{\alpha}}} \biggr\}
^{T}
\\
&&\hspace*{140pt}{}\times \operatorname{Vec} \bigl( \bolds{\psi }_{j}(t_{jl})\bolds{\psi}_{j}(t_{jl})^{\prime} \bigr) \biggr],
\end{eqnarray*}
with $\frac{\partial^{2}\mathbf{J}(\bolds{\alpha},\bolds{\theta},%
\bolds{\lambda})}{\partial\bolds{\alpha}_{j}^{\prime}\,\partial
\bolds{\alpha}_{j}} $ and $\frac{\partial^{3}\mathbf{J}(\bolds{\alpha},\bolds{\theta},%
\bolds{\lambda})}{\partial\lambda_{k}\,\partial\bolds{\alpha}_{j}^{\prime}\,\partial\bolds{\alpha}_{j}}$ also given in the \hyperref[app]{Appendix}.

For generalized profile estimation, it is standard to use B-spline bases
with equally spaced knots. By analogy with penalized splines, it is expected
that a sufficiently large number of basis vectors $K$ (e.g., $n/2$,
where $n$
is the number of data points) would be appropriate to minimize the bias
[\citet{Rametal07}]. However, the results of \citet{ClaKriOps09} imply
that with a large number of equally spaced knots, penalty weight is the only
effective smoothing parameter for penalized splines. Thus, if nonsmooth
input functions are present in the ODE model, then penalty (\ref{PEN}) may
not serve the \mbox{purpose} of regularization as a roughness penalty. In
particular, this is the case in our data with insulin infusion rate
being a
piecewise constant step function. Therefore, it is proposed to optimize the
number and location of knots for B-spline bases. With a small number of
knots, penalized splines behave asymptotically as least-squares splines and
the knots become operative smoothing parameters [\citet{ClaKriOps09}].
Optimizing the number and/or location of knots has been previously
considered in the context of free-knot splines [e.g., \citet{Jup78}; \citet{Hu93};
\citet{Lin99}]. It may be performed using the standard model \mbox{selection}
criteria such as the Akaike information criterion (AIC), the Bayesian
information
criterion (BIC) and GCV, but it is known to be very computationally
intensive and often intractable using derivative-based optimization
methodology. \mbox{Recently,} \citet{Spietal13} developed two stochastic search
algorithms for selecting both the number and location of knots. The
algorithms combined with adjusted GCV, AIC or BIC optimization criteria are
implemented in the R package \texttt{freeknotsplines} available at the
Comprehensive R Archive Network (CRAN) \mbox{(\url{http://cran.r-project.org/})}. \citet{Spietal13} report excellent numerical performance of the new
algorithms at
producing knot locations that are near optimal in the sense of average
squared error loss. We propose to adapt this state-of-the-art methodology
for selection of free knots for B-spline bases representing the ODE
solution(s) for generalized smoothing. Knots selection is optimized using
only the observed data, without taking into account the ODE model. This
allows using the standard penalization of the first or second
derivative for
the purpose of selecting the knots for penalized spline optimally smoothing
the data. Then the optimized sequence of knots is used to construct
the corresponding B-spline basis for representing the ODE solution for
generalized profiling estimation, which can be performed using the R package
\texttt{CollocInfer} or \mbox{Matlab} software functions for profiled estimation
of differential equations available at
\url{http://faculty.bscb.cornell.edu/~hooker/profile\_webpages/}. To
obtain the starting values and weights for the error sum-of-squares required
for implementation of the generalized profile estimation, we first smooth
the data points nonparametrically and optimally in the classic sense of
cross-validation and then use the standard nonlinear regression methods to
estimate the structural parameters of ODEs. The resulting estimates
serve as
the starting values for the generalized profile estimation. The necessary
weights are the reciprocals of the \mbox{cross-}validation estimates for the
standard deviation of the errors, $w_{j}=\hat{\sigma}_{j}^{-2}$.

\section{Application to glucose--insulin dynamics in T1DM subjects}\label{sec5}

For parameter estimation, model (\ref{mainODE}) was reparameterized as
the following to impose physiologically meaningful signs of the original
parameters:%
\begin{eqnarray}\label{reparODE}
\overset{\bullet} {G}(t) &=&\theta_{1}-e^{\theta_{2}}G(t)-e^{\theta
_{3}}G(t)I(t)\nonumber
\\
&&{}+\llvert \theta_{6}\rrvert (t-t_{M_{1}})_{+}e^{-\llvert
\theta_{7}\rrvert (t-t_{M_{1}})_{+}}
\nonumber\\[-8pt]\\[-8pt]
&&{}+\llvert \theta_{8}\rrvert (t-t_{M_{2}})_{+}e^{-\llvert \theta_{9}\rrvert (t-t_{M_{2}})_{+}},\nonumber
\\
\overset{\bullet} {I}(t) &=&-e^{\theta_{4}}I(t)+\theta_{5}r(t),
\nonumber
\end{eqnarray}
so that $b_{0}=\theta_{1}, b_{1}=e^{\theta_{2}}, b_{2}=e^{\theta
_{3}}$, $c_{1}=e^{\theta_{4}}, c_{2}=\theta_{5}, \mu_{1}=\llvert
\theta_{6}\rrvert, \nu_{1}=-\llvert \theta_{7}\rrvert,
\mu_{2}=\llvert \theta_{8}\rrvert, \nu_{2}=-\llvert
\theta
_{9}\rrvert$, and a $9\times1$ vector of all structural
parameters is $%
\bolds{\theta}= [ \theta_{1},\ldots,\theta_{9} ] $.
System (\ref{reparODE}) may be written in the general form (\ref{ssODE})
with $\mathbf{x}(t)= [ x_{1}(t),x_{2}(t) ] ^{\prime}$ and
\begin{eqnarray*}
f_{1} \bigl( \mathbf{x}(t),\mathbf{u}(t),t,\bolds{\theta} \bigr) &=&
\theta _{1}-e^{\theta_{2}}x_{1}(t)-e^{\theta_{3}}x_{1}(t)x_{2}(t)
\\
&&{}+\llvert \theta_{6}\rrvert u_{1}(t)\exp \bigl\{ -\llvert
\theta _{7}\rrvert u_{1}(t) \bigr\} +\llvert
\theta_{8}\rrvert u_{2}(t)\exp \bigl\{ -\llvert
\theta_{9}\rrvert u_{2}(t) \bigr\},
\\
f_{2} \bigl( \mathbf{x}(t),\mathbf{u}(t),t,\bolds{\theta} \bigr)
&=&-e^{\theta_{4}}x_{2}(t)+\theta_{5}u_{3}(t),
\end{eqnarray*}
where $x_{1}(t)=G(t)$, $x_{2}(t)=I(t)$, $u_{1}(t)=(t-t_{M_{1}})_{+}$, $u_{2}(t)=(t-t_{M_{2}})_{+}$, $u_{3}(t)=r(t)$, $\mathbf{u}(t)= [
u_{1}(t),u_{2}(t),u_{3}(t) ] ^{\prime}$.

\begin{figure} 

\includegraphics{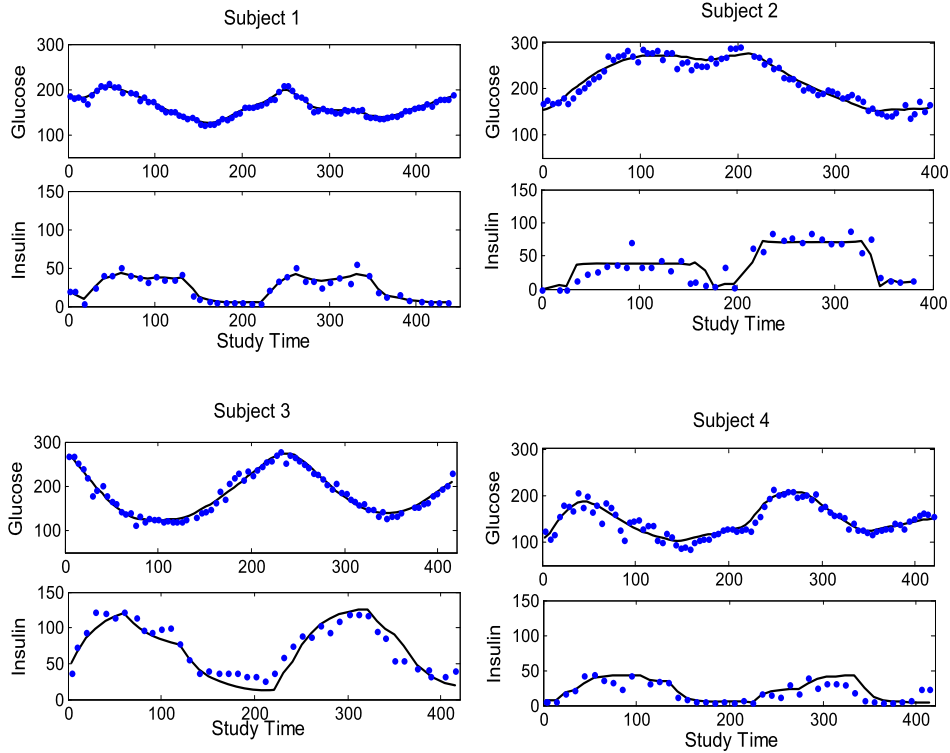}

\caption{Fitted ODE solutions for 4 T1DM subjects.}\label{Fig2}
\end{figure}

The number and locations of knots were optimized separately for glucose and
insulin in each T1DM subject using the corrected Akaike information
criterion (AICc) as implemented in R package \texttt{freeknotsplines}. The
resulting numbers of optimized knots ranged between 12 and 18 for the
glucose component and between 6 and 13 for the insulin component. Using
these small numbers of knots resulted in good nonparametric smoothing of
observed values, but ODE parameter estimates were not as physiologically
plausible as using 30--40 equally spaced knots. Therefore, for each subject,
the sets of optimal nodes for the glucose and insulin components were pooled
into one set of 15--23 knots (dropping one of the knots from resulting pairs
of knots with the distance of less than 5 min). Figure~\ref{Fig2} shows the resulting
B-spline approximations of the ODE solutions for all subjects. It is clear
that the approximate ODE solutions track well the global trends in the
observed data. Parameter estimates from the fitted ODE model were used to
compute most important physiologically meaningful quantities, for which the
ranges were previously described in the diabetes research literature. This
serves as an additional validation step for the models fitted to the
real data.
The metabolic clearance rate (MCR) [(min$^{-1}$)~ml/kg] is the product
of the
fractional clearance rate and distribution volume in ml of insulin
normalized by the subject's weight. The fractional clearance rate and
distribution volume are $c_{1}$ and $1000c_{2}^{-1}$, respectively,
where $%
c_{1}$ and $c_{2}$ are parameters in (\ref{Ht}). Thus, the metabolic
clearance rate is computed as $\mathrm{MCR}=1000c_{2}^{-1}c_{1}/\mathrm{weight}$. The basal
insulin infusion rate ($r_{b}$) required to maintain euglycemia was
calculated to investigate the physiological validity of the parameters in
equation (\ref{mainODE}). By setting to 0 the glucose derivative and glucose
appearance from a meal (steady-state conditions), and setting the glucose
concentration to the euglycemic value of $G_{b}=80$ [mg/dl] in
(\ref{mainODE}), it is possible to compute the basal insulin value
required to
achieve a euglycemic value of 80 mg/dl as $I_{b}= (
b_{0}-b_{1}G_{b} ) /b_{2}G_{b}$. Then the infusion rate required to
achieve $I_{b}$ is computed from (\ref{Ht}) by setting $\overset
{\bullet}{I}(t)=0$ ($r_{b}=c_{2}^{-1}c_{1}I_{b})$. Table~\ref{tab1} presents
the estimates of
parameters in model (\ref{mainODE}), computed from the optimized generalized
profiling estimates of structural parameters in (\ref{reparODE}), subject
weights and computed MCR and~$r_{b}$. The estimated values of MCR fall
within the range [\mbox{7.5--35.2 (min$^{-1}$)~ml/kg}] previously reported in the
literature [\citet{Tho90}]. The computed $r_{b}$ values range from
0.22 to 0.34 U/hr (3.7 to 5.7 mU/min). These estimates are feasible for T1DM
patients with low to normal basal insulin requirements [\citet{SchBoy05}].

%

%
\begin{table}[b]
\tabcolsep=0pt
\caption{The estimates of the ODE parameters and corresponding
metabolic clearance rate (MCR) and basal insulin infusion rate
($r_{b}$) for all T1DM subjects}\label{tab1}
\begin{tabular*}{\tablewidth}{@{\extracolsep{\fill}}@{}lcccccccc@{}}
\hline
\textbf{Subject} & $\bolds{b_{0}}$ & $\bolds{b_{1}}$ & $\bolds{b_{2}}$ & $\bolds{c_{1}}$ & $\bolds{c_{2}}$ & \textbf{Weight} & \textbf{MCR} & $\bolds{r_{b}}$\\
\hline
1 & 0.95 & 0.001 & 0.0002 & 0.05 & 0.04 & 73.5 & 20.9 & 0.30 \\
2 & 0.46 & 0.001 & 0.0001 & 0.25 & 0.17 & 65.8 & 23.2 & 0.29 \\
3 & 1.94 & 0.001 & 0.0002 & 0.03 & 0.04 & 61.3 & 13.4 & 0.34 \\
4 & 1.24 & 0.005 & 0.0003 & 0.11 & 0.06 & 74.9 & 23.5 & 0.22 \\
\hline
\end{tabular*}
\end{table}

%
\begin{figure} 

\includegraphics{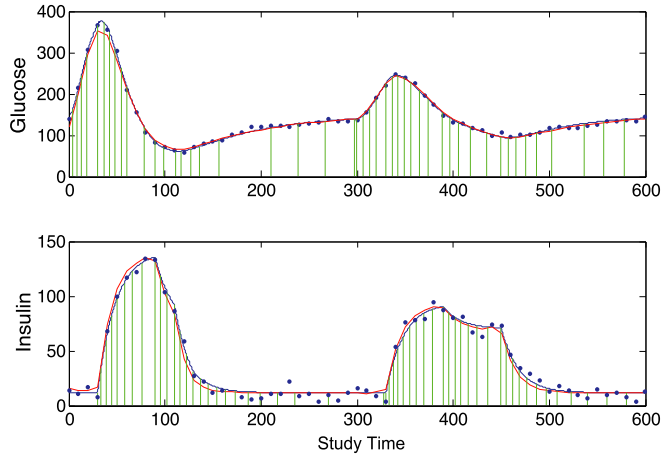}

\caption{Sample simulated data (blue points) with the true ODE solution
(blue line) and generalized profile solution (red line). Locations of
optimized knots for B-spline basis shown as vertical green lines (47
knots for glucose and 54 knots for insulin).}\label{Fig3}
\end{figure}

%
\begin{figure} 

\includegraphics{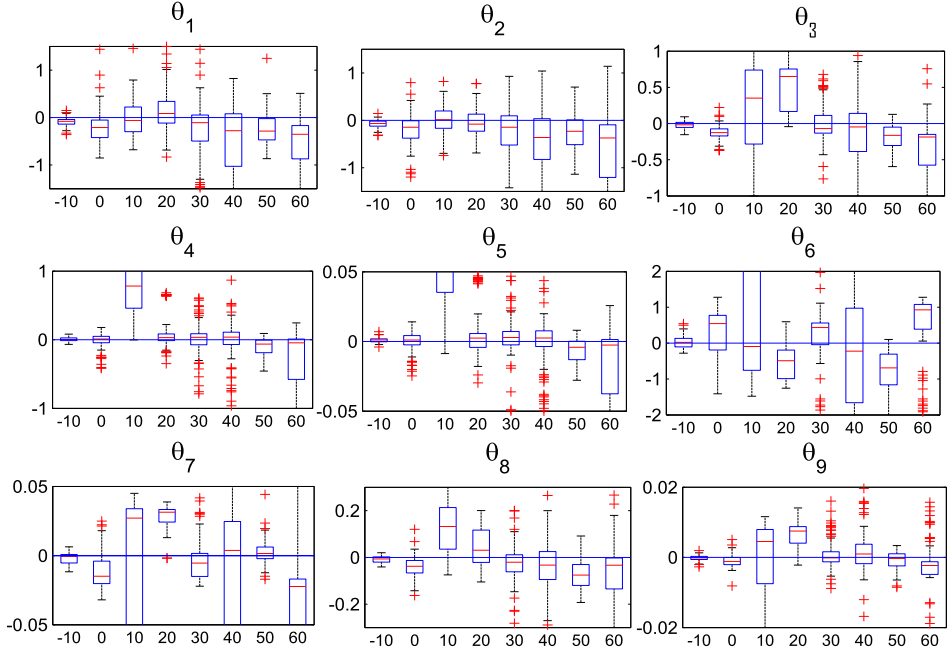}

\caption{Error distribution of ODE parameter estimates. The errors
corresponding to ODE solutions represented by B-splines with fixed a
priori chosen number K of basis functions ($K={}$10, 20, 30, 40, 50, 60)
are shown with $x$-coordinate equal to $K$. The errors corresponding to ODE
solutions represented by B-splines with equally spaced knots and the
number of basis functions K that minimizes the outer optimization
criterion over the grid {10, 20, 30, 40, 50, 60} are shown with
$x$-coordinate zero. The errors corresponding to ODE solutions using
B-splines with optimized number and location of knots are shown with
$x$-coordinate $-$10.}\label{Fig4}
\end{figure}

\section{Simulation study}\label{sec6}

A small simulation study was conducted to evaluate performance of the
proposed optimization for a nonlinear and nonhomogeneous ODE model (\ref
{mainODE}). The input functions used in the model were \mbox{generated} separately
according to the study protocol for IV insulin and Monks' (1990) model of
glucose appearance in the blood. Then exact numeric solutions were computed
for the postulated ODE model (\ref{mainODE}) with known parameters
$\bolds{\theta}\,{=}\, [2.08,-4.60,-7.99,-2.91,0.08,2.15,-0.07, 0.39,-0.03 ] $
(similar to parameters estimated from real data) and input functions.
Finally, 100 data sets with $N=61$ observations each (corresponds to
sampling every 10 minutes during 6 hours) were generated according to the
measurement error model with exact numeric ODE solution as the mean
function and independent Gaussian errors with mean zero and variance $25$.
The optimized generalized profiling estimation was performed as
described in
Sections~\ref{sec3}--\ref{sec4} using B-spline bases with (i) equally spaced knots a priori
chosen from the set $\bolds{\Omega}= \{
10,20,30,40,50,60 \}$, (ii) equally spaced knots with the number of
basis functions $K$ that minimizes (\ref{outerF}) over $K\in
\bolds{\Omega}$, and (iii) optimized number and location of knots
in the range 5--60
as described in Section~\ref{sec4}. Option (ii) corresponds to optimizing the number
of equally spaced knots using a sparse full-search algorithm similar to the
one described in \citet{Rup02}. Figure~\ref{Fig3} shows a sample simulated data set
(blue points) with the true ODE solution (blue line), generalized profile
solution (red line) and locations of optimized knots (47 knots for glucose
and 54 knots for insulin with locations shown by vertical green lines). As
expected for optimized knot selection, more knots are placed in the
intervals of rapid change of output functions as compared to the intervals
with small gradients.

\begin{figure} 

\includegraphics{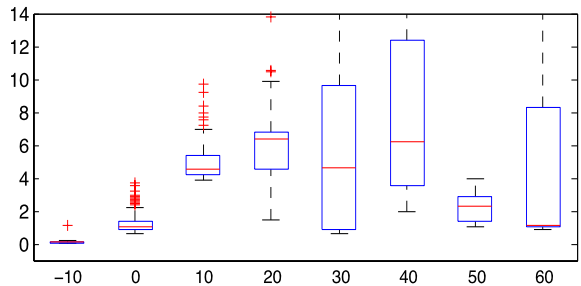}

\caption{Distribution of root mean prediction error of ODE solutions
estimated on the same grid as original data points. The prediction
errors of ODE solutions represented by B-splines with fixed a priori
chosen number K of basis functions ($K={}$10, 20, 30, 40, 50, 60) are shown
with $x$-coordinate equal to $K$. The prediction errors of ODE solutions
represented by B-splines with equally spaced knots and the number of
basis functions $K$ that minimizes the outer optimization criterion over
the grid {10, 20, 30, 40, 50, 60} are shown with $x$-coordinate zero. The
prediction errors of ODE solutions using B-splines with optimized
number and location of knots are shown with $x$-coordinate $-$10.}\label{Fig5}
\end{figure}

Figure~\ref{Fig4} shows the error distributions of the resulting ODE parameter
estimates. The errors corresponding to ODE solutions represented by
B-splines with fixed a priori chosen number $K\in\bolds{\Omega}$
of basis functions are shown with the $x$-coordinate equal to $K$. The errors
corresponding to equally spaced knots\vadjust{\goodbreak} and $K$ minimizing~(\ref{outerF}) over
$K\in\bolds{\Omega}$ are shown with $x$-coordinate zero. The
errors corresponding to optimized number and location of knots are shown
with $x$-coordinate $-$10. None of the a priori chosen values of $K$ yields
accuracy of estimating ODE parameters comparable to the accuracy achieved
using equally spaced knots with $K$ minimizing (\ref{outerF}) over $K\in
\bolds{\Omega}$ or optimized unequally spaced knots. For
each $%
\theta_{i}$, some values of a priori chosen $K$ yielded similar error
distributions, however, such values of $K$ are different for different ODE
parameters, which does not allow selecting a common fixed optimal $K$. For
majority of $\theta_{i}$, the optimized generalized profiling with
unequally spaced knots yields the smallest estimation errors. Figure~\ref{Fig5} shows
the distribution of root mean prediction errors of approximated ODE
solutions. For each simulated data set, the mean prediction error was
computed on the same grid as original data points using the true values
$%
x_{j}(t_{jl})$, which are known for the simulated data.  Optimizing both
the number and location of knots provides dramatic reduction in the root
mean prediction error of ODE solutions as compared to using an a priori
chosen $K$ and substantial reduction as compared to optimizing just the
number of equally spaced knots. Also, optimizing just the number of equally
spaced knots provides large reduction in root mean prediction error of ODE
solution as compared to using an a priori chosen $K$.

In conclusion, optimizing the number and location of knots for B-spline
approximating the ODE solution provides the most accurate estimates of the
ODE parameters and solution. The improvement well justifies additional
computational cost for such optimization. Meanwhile, computational
costs of
repeated optimized generalized smoothing for multiple numbers of equally
spaced knots are substantially higher than optimizing the number and
location of knots once.

%

\section{Discussion}\label{sec7}

In this work we proposed an approach for optimizing the generalized
profiling estimation of models defined by nonlinear and/or nonhomogeneous
ODE systems. Our approach includes optimization with respect to the penalty
parameters and with respect to the number and location of knots for \mbox{B-}spline
basis used to approximate the ODE solution. Additional regularization from
optimizing the knots for the \mbox{B-}spline basis is especially important in
the case
of discontinuous input functions. The covariance penalties criterion for
outer optimization of penalty weights is equivalent to the generalized
cross-validation criterion in the case of linear prediction rules, but
it is
applicable to a much wider range of nonlinear and/or nonhomogeneous ODE
models that result in nonlinear prediction rules.

Applying the optimized generalized profiling to glucose and insulin
concentration data in T1DM patients, we obtained physiologically plausible
results for the proposed parsimonious model of glucose--insulin dynamics.
Thus, unlike the majority of existing mathematical models of glucose--insulin
dynamics, our model is estimable from a relatively small number of noisy
observations of glucose and insulin concentrations.
Our approach renders the proposed model an attractive candidate
for developing automated algorithms of IV insulin delivery for
in-hospital glucose management of TIDM patients, as well as for
other scenarios involving medical management of chronic conditions.

\begin{appendix}
\section*{Appendix: Derivatives}\label{app}
Assuming that it is appropriate to integrate under the integral in the
penalty function, omitting the limits of integration and dropping the
explicit dependence on $t$, derivatives for Gauss--Newton
optimization are
\begin{eqnarray*}
\frac{\partial\mathbf{J}(\bolds{\alpha}|\bolds{\theta},\bolds{\lambda
})}{\partial\bolds{\alpha}_{j}} &=&-2w_{j}\bolds{\Psi}_{j}^{\prime
}
( \mathbf{y}_{j}-\bolds{\Psi}_{j}\bolds{\alpha}_{j} )
\\
&&{}+\sum_{k=1}^{d}2\lambda_{k}
\int \bigl( {\overset{\bullet}{\bolds{\psi}}}{}_{k}^{\prime}
\bolds{\alpha}_{k}-f_{k} ( \tilde {
\mathbf{x}},\mathbf{u},s,\bolds{\theta} ) \bigr) \biggl(
\delta_{kj}{\overset {\bullet}{\bolds{\psi}}}_{k}-
\frac{\partial f_{k} ( \tilde
{\mathbf{x%
}},\mathbf{u},s,\bolds{\theta} ) }{\partial\bolds{\alpha}_{j}} \biggr) \,ds,
\end{eqnarray*}
where $\delta_{kj}=1$, if $k=j$, and $\delta_{kj}=0$, if $k\neq j$.  Then,
\begin{eqnarray*}
\frac{\partial\mathbf{J}(\bolds{\alpha}|\bolds{\theta},\bolds{\lambda
})}{\partial\lambda_{k}\,\partial\bolds{\alpha}_{j}}
&=& 2\int \bigl( {\overset{\bullet} {\bolds{\psi}}}{}_{k}^{\prime}
\bolds{\alpha}_{k}-f_{k} ( \tilde{\mathbf{x}},
\mathbf{u},s,\bolds{\theta} ) \bigr) \biggl( \delta_{kj}{\overset{
\bullet} {\bolds{\psi}}}_{k}-\frac{\partial
f_{k} ( \tilde{\mathbf{x}},\mathbf{u},s,\bolds{\theta} )
}{\partial\bolds{\alpha}_{j}} \biggr) \,ds,
\\
\frac{\partial^{2}\mathbf{J}(\bolds{\alpha}|\bolds{\theta},\bolds{\sigma},
\bolds{\lambda})}{\partial\bolds{\alpha}_{i}'\,\partial
\bolds{\alpha}_{j}}
&=&\delta_{ij}2w_{j}\bolds{\Psi}_{j}^{\prime
}\bolds{\Psi}_{j}
\\
&& {}+\sum_{k=1}^{d}2\lambda_{k}
\int \biggl( \delta_{ki}{\overset{\bullet } {\bolds{\psi}}}_{k}-\frac{\partial f_{k} ( \tilde{\mathbf{x}},
\mathbf{u},s,\bolds{\theta} ) }{\partial\bolds{\alpha}_{i}} \biggr)
\\
&&\hspace*{55pt}{}\times \biggl( \delta_{kj}
{\overset{\bullet} {\bolds{\psi}}}_{k}-\frac{\partial
f_{k} ( \tilde{\mathbf{x}},\mathbf{u},s,\bolds{\theta} )
}{\partial\bolds{\alpha}_{j}} \biggr)
^{\prime}\,ds
\\
&&{}+\sum_{k=1}^{d}2\lambda_{k}
\int \bigl( {\overset{\bullet} {\bolds{\psi}}}{}_{k}^{\prime}
\bolds{\alpha}_{k}-f_{k} ( \tilde {
\mathbf{x}},%
\mathbf{u},s,\bolds{\theta} ) \bigr) \biggl( -
\frac{\partial
^{2}f_{k} ( \tilde{\mathbf{x}},\mathbf{u},s,\bolds{\theta} )
}{\partial\bolds{\alpha}_{i}'\,\partial\bolds{\alpha}_{j}} \biggr) \,ds,
\\
\frac{\partial^{3}\mathbf{J}(\bolds{\alpha}_{j}|\bolds{\theta},\bolds{\sigma},\bolds{\lambda})}{\partial\lambda_{k}\,\partial
\bolds{\alpha}_{i}'\,\partial\bolds{\alpha}_{j}}
&=&2\int \biggl( \delta_{ki}{\overset{\bullet} {
\bolds{\psi}}}_{k}-\frac
{\partial f_{k} ( \tilde{\mathbf{x}},\mathbf{u},s,\bolds{\theta} ) }{\partial\bolds{\alpha}_{i}} \biggr) \biggl( \delta
_{kj}{\overset{\bullet} {\bolds{\psi}}}_{k}-
\frac{\partial f_{k} (
\tilde{\mathbf{x}},\mathbf{u},s,\bolds{\theta})}{\partial\bolds{\alpha}_{j}}\biggr)^{\prime}\,ds
\\
&&{}+2\int \bigl( {\overset{\bullet} {\bolds{\psi}}}{}_{k}^{\prime}
\bolds{\alpha}_{k}-f_{k} ( \tilde{\mathbf{x}},
\mathbf{u},s,\bolds{\theta} ) \bigr) \biggl( -\frac{\partial^{2}f_{k} (
\tilde{\mathbf{x}},\mathbf{u},s,\bolds{\theta} ) }{\partial
\bolds{\alpha}_{i}'\,\partial\bolds{\alpha}_{j}} \biggr) \,ds.
\end{eqnarray*}
\end{appendix}

\section*{Acknowledgments}
The authors are grateful to the Editor and reviewers for constructive
comments that helped to improve the quality and accessibility of this
manuscript.




\printaddresses

\end{document}